\documentclass[final]{IEEEtran}

\usepackage{multirow}
\usepackage{latexsym}
\usepackage{graphicx}
\usepackage{float}
\usepackage{amsmath}
\usepackage{amsthm}
\usepackage{lipsum}
\usepackage{subfig}
\usepackage{graphicx}
\usepackage{authblk}
\usepackage{bm}
\usepackage{booktabs}
\usepackage{amsthm}
\usepackage[section]{placeins}
\usepackage{soul}

\usepackage{colortbl}

\usepackage{enumitem}

\usepackage{CJKutf8}

\newtheorem*{remark}{Remark}

\newtheorem{lemma}{Lemma}
\newtheorem{example}{Example}
\newtheorem{definition}{Definition}

\makeatletter  
\newif\if@restonecol  
\makeatother

\usepackage[linesnumbered,ruled,vlined]{algorithm2e}
\usepackage{algpseudocode}  
\usepackage{amsmath}

\usepackage{amssymb}
\DeclareMathOperator*{\argmax}{argmax}

\usepackage{mathrsfs}
\usepackage{subfig}
\usepackage{caption}
\newcounter{problem}

\SetKwInput{KwInitialize}{Initialize}

\renewcommand{\baselinestretch}{1.0}

\begin{document}

\title{Joint Resource Estimation and Trajectory Optimization for eVTOL-involved CR network: A Monte Carlo Tree Search-based Approach}


\author{Kai Xiong,~\IEEEmembership{Member,~IEEE}, Chenxin Yang, Yujie Qin,~\IEEEmembership{Member,~IEEE}, Wanzhi Ma, Chau Yuen,~\IEEEmembership{Fellow,~IEEE}

\thanks{
%

K. Xiong, C. Yang, Y. Qin, and Wanzhi Ma are with School of Information and Communication Engineering, University of Electronic Science and Technology of China (UESTC), Chengdu, 611731, China; and, Shenzhen Institute for Advanced Study, UESTC, Shenzhen, 518110, China.
}


\thanks{
C. Yuen is with the School of Electrical and Electronic Engineering, Nanyang Technological University, Singapore. 
}




\thanks{The financial support of National Natural Science Foundation of China (NSFC), Grant No.62201122.
}


\thanks{The corresponding author is Wanzhi Ma, email: mawanzhi@uestc.edu.cn}
}


\maketitle


\begin{abstract}
Electric Vertical Take-Off and Landing (eVTOL) aircraft, pivotal to Advanced Air Mobility (AAM), are emerging as a transformative transportation paradigm with the potential to redefine urban and regional mobility. While these systems offer unprecedented efficiency in transporting people and goods, they rely heavily on computation capability, safety-critical operations such as real-time navigation, environmental sensing, and trajectory tracking—necessitating robust offboard computational support.
A widely adopted solution involves offloading these tasks to terrestrial base stations (BSs) along the flight path. However, air-to-ground connectivity is often constrained by spectrum conflicts with terrestrial users, which poses a significant challenge to maintaining reliable task execution. Cognitive radio (CR) techniques offer promising capabilities for dynamic spectrum access, making them a natural fit for addressing this issue.
Existing studies often overlook the time-varying nature of BS resources, such as spectrum availability and CPU cycles, which leads to inaccurate trajectory planning, suboptimal offloading success rates, excessive energy consumption, and operational delays.
To address these challenges, we propose a trajectory optimization framework for eVTOL swarms that maximizes task offloading success probability while minimizing both energy consumption and resource competition (e.g., spectrum and CPU cycles) with primary terrestrial users. The proposed algorithm integrates a Multi-Armed Bandit (MAB) model to dynamically estimate BS resource availability and a Monte Carlo Tree Search (MCTS) algorithm to determine optimal offloading decisions, selecting both the BSs and access time windows that align with energy and temporal constraints. Furthermore, the framework incorporates adaptive trajectory re-planning to account for BS access failures during execution.
Simulation results reveal that the proposed approach outperforms benchmark methods, achieving up to a 60\% improvement in energy efficiency and a 17\% increase in task completion probability, demonstrating its suitability for large-scale AAM deployments.

\end{abstract}


\IEEEpeerreviewmaketitle

\section{Introduction}

\IEEEPARstart {A}{dvanced} Air Mobility (AAM) refers to a three-dimensional (3D) aerial transportation system that incorporates next-generation technologies such as remote pilote, autonomous, and electric Vertical Take-Off and Landing (eVTOL) aircraft. 
These systems are expected to revolutionize the movement of people and goods by enabling on-demand, airborne transportation that operates independently of ground infrastructure \cite{johnson2022nasa,guo2024advanced,cohen2021urban,goyal2021advanced}.
AAM and eVTOL offer efficient, eco-friendly \cite{aam1}, and wide-coverage transportation solutions, ranging from urban centers to rural areas \cite{10959008Bang}. Additionally, aviation authorities are actively exploring the integration of advanced Communication, Computing, and Control (3C) technologies to enable intelligent and collaborative eVTOL operations in low-altitude environments \cite{ICAAMap2021}.


While intelligent eVTOL applications hold great potentials, they also face several critical challenges. One key issue is the heavy computational demand required to support safety-critical functions such as safe navigation, target sensing, and tracking \cite{10388419}. Due to the limited processing capabilities of onboard chips, eVTOLs often struggle to handle these heavy-computational tasks effectively. 
This limitation underscores the importance of task offloading, whereby computational tasks are offloaded to nearby ground base stations (BSs) that offer greater processing power and enhanced quality-of-service (QoS).
Typically, the effectiveness of task offloading depends on reliable air-to-ground (A2G) communication and the availability of BS resources, such as idle spectrum and CPU cycles. It is also influenced by the distance between eVTOLs and BSs, as well as the trajectories of the eVTOLs during their missions.

In general, ground BSs are primarily deployed to serve terrestrial users, commonly referred to as primary users in the Cognitive Radio (CR) context. As a result, they may not be able to efficiently support secondary users \cite{chen2022dedicating,qin2023coverage}, such as aerial users like eVTOLs. This can lead to task offloading failures, an issue that is often overlooked in the existing literature. For example, the authors in \cite{9181432Kai} focus on maximizing task offloading allocations without accounting for potential access conflicts between primary and secondary users.
Another challenge arises from the limited battery capacity of eVTOLs, which restricts their ability to sustain long-duration tasks \cite{granado2022machine,warren2019effects,qin2020performance}. 
{\color{black} Consequently, the task offloading process must carefully consider three critical factors: offloading duration, eVTOL trajectories, and energy consumption. This comprehensive consideration is essential to ensure successful task execution while meeting both operational and energy constraints.
}

The aforementioned two issues are interrelated. On one hand, the selection of BSs for task offloading directly influences the eVTOL's flight trajectory and corresponding energy consumption. On the other hand, if an eVTOL fails to access the visited BS, it not only results in wasted onboard energy but also necessitates additional energy to search for alternative BSs and restart the task offloading process. Therefore, determining the optimal flight trajectory, BS selection, and task offloading strategy under both energy and task constraints is a non-trivial challenge for eVTOLs.

{\color{black}
To address these challenges, we develop a CR-enabled approach for BS spectrum access that addresses these constraints. Our method jointly designs an optimized trajectory algorithm for eVTOL operations. This integrated solution maximizes the probability of successful task offloading while efficiently managing spectrum resources and energy consumption.
}

\subsection{Contributions}
Existing research on these challenges, particularly within the context of cognitive radio-based trajectory design for eVTOLs, remains limited. To address these issues and enhance the ability of eVTOLs to successfully complete data transmission tasks, this work proposes a Monte Carlo Tree Search (MCTS)-based BS spectrum access scheduling scheme that jointly considers BS selection, access periods, and flight energy consumption. 
The main contributions of this work are summarized as follows:

\begin{itemize}
\item We propose an MCTS-based decision-making scheme that jointly determines BS selection and offloading period for eVTOLs. More importantly, in practical scenarios, the proposed scheme can effectively handle task offloading failures by enabling the eVTOL to dynamically re-evaluate and update its decisions. This allows the system to adapt to unexpected conditions and complete tasks seamlessly.

\item The proposed scheme not only facilitates task offloading decisions for eVTOLs but enables spatio-temporal planning for the aviation routes while minimizing flight energy consumption. Furthermore, the scheme supports both offline and online implementation, it can be used to pre-plan flight schedules prior to takeoff and to dynamically adjust the flight route in real time based on environmental changes and BS conditions.

\item Compared to traditional algorithms, the proposed approach demonstrates superior performance by achieving faster convergence and more accurate results. It enables the eVTOL to obtain information about BS resource status across different time periods through the proposed prior estimation method, facilitating accurate decision-making and significantly reducing the flight energy cost from the algorithm's trial-and-errors.
\end{itemize}

The remainder of this paper is organized as follows. Section II reviews the related work. Section III presents the system model, introduces the energy consumption model, BS resource availability model, and trajectory design objective functions. Section IV provides a detailed performance analysis, solves the objective functions, and proposes the proposed MCTS-based decision-making scheme. Simulation results are presented in Section V. Finally, conclusions are drawn in Section VI.




\section{Related Work}

Literature related to this work can be categorized into: (i) Trajectory design of AAM, (ii) CR-enabled eVTOL/UAV\footnote{Here, we also include UAV-related works, as UAVs are considered a subset of eVTOL aircraft.} system, and (iii) eVTOL/UAV-based task offloading analysis. A brief discussion on related works in each of these categories is discussed in the following lines.

\subsection{Trajectory Design of AAM}
The authors in \cite{aam1} outlined critical technologies for low-altitude networks, air traffic management, and AAM system security frameworks, while the authors in \cite{10388419} provided a comprehensive review of autonomous navigation for eVTOLs. In \cite{aam3}, the authors optimized UAV mobility, communication, and computing offloading in A2G networks, proposing a gradient projection-based algorithm to minimize UAV energy consumption. A tilt-wing eVTOL takeoff trajectory optimization problem was analyzed in \cite{chauhan2020tilt}, where the authors demonstrated the impact of wing loading and available power on optimized flight paths. In \cite{pradeep2020wind}, the authors compared wind-optimal and great-circle trajectories for short urban flights, finding minimal operational advantages in terms of energy consumption and flight duration, though wind conditions significantly influenced the results. To further optimize energy and aviation efficiency in eVTOL flights through dynamic wind fields, \cite{liu2024teevtol} introduced a novel path planning method that leverages deep reinforcement learning to determine the shortest paths in a weighted graph, where the weights represent either energy or time costs. Additionally, several energy-efficient trajectory optimization methods for the landing and takeoff phases were proposed in \cite{wang2021energy, park2023trajectory, yeh2023inverse}.
{\color{black}
Nevertheless, a critical research gap persists in the integrated optimization of eVTOL task offloading, CR spectrum \& computing access, and energy-efficient trajectory design for comprehensive AAM operations.
}

\subsection{CR-enabled eVTOL/UAV System}
In \cite{cr1}, the authors proposed a framework for cognitive UAVs (CUAVs) that dynamically adjusts flight paths and transmission power to avoid disrupting primary users. In \cite{nobar2021resource}, the performance of a cognitive radio-enabled UAV network configuration was investigated, where UAVs communicate with secondary ground terminals in a licensed spectrum band. A 3D trajectory and resource optimization for a UAV-relay-assisted CR network was analyzed in \cite{9459574}, where the authors introduced an algorithm to solve the challenging non-convex problem using the path discretization technique. In \cite{9714863}, the authors examined the average network throughput in a UAV-enabled CR network, considering the UAV as a secondary user. They exploited the joint design of UAV trajectory and resource allocation to maximize average throughput while adhering to constraints on co-channel interference and completion time. In \cite{9867917}, a CUAV-assisted network was analyzed, where the authors jointly optimized the UAV trajectory and communication to balance data collection and transmission, achieving high offloading efficiency. Furthermore, \cite{deng2021joint} investigated joint UAV trajectory and power allocation optimization for the Non-Orthogonal Multiple Access (NOMA) protocol in cognitive radio networks, while \cite{nguyen2021uav} explored UAV-assisted secure communication in terrestrial CR networks.
{\color{black}
Despite these contributions to CR-enabled UAV systems, current literature has yet to adequately consider the heavy task offloading requirements to CR-enabled aerial traffic scenarios.
}

\subsection{UAV-based Task Offloading Analysis}
In \cite{uav1}, the authors proposed a digital twin (DT)-assisted deep reinforcement learning (DRL) framework to optimize computation offloading in UAV-MEC systems, minimizing task processing delay by jointly optimizing flight paths and offloading ratios. In \cite{uav2}, the authors studied UAV-assisted target tracking in maritime scenarios, offloading image processing tasks to unmanned surface vehicles (USVs) to address the UAV's energy and computation limitations. Additionally, in \cite{uav3}, the authors introduced DroneSegNet, a semantic segmentation framework for aerial images, offloading computationally intensive deep learning tasks to the cloud.
In \cite{uav5}, the authors proposed an energy-efficient transmission strategy for UAV patrol inspection systems, using a weighted factor to balance flight distance and offloading efficiency. In satellite IoT networks, \cite{uav4} modeled multi-task offloading as a directed acyclic graph (DAG) and proposed an attention mechanism-proximal policy optimization (A-PPO) algorithm to manage task dependencies. Furthermore, \cite{uav6} addressed dynamic task offloading in multi-UAV vehicular edge computing (UVEC), employing stochastic network calculus (SNC) to derive delay and buffer constraints.
In intelligent reflecting surfaces (IRS)-assisted edge computing systems, \cite{uav8} proposed a two-stage deep energy optimization approach, combining deep learning with IRS to enhance signal propagation and reduce energy consumption during offloading. Both \cite{uav2} and \cite{uav4} addressed energy efficiency by offloading computation to more powerful edge nodes, such as satellites and USVs, demonstrating that hierarchical offloading (local UAV processing combined with edge/cloud assistance) can effectively balance latency and energy usage.
{\color{black}
However, current literature has largely overlooked the intricate interplay between spectrum availability and computational offloading efficiency, representing a crucial frontier for future research in UAV edge computing systems.

\begin{table}[h]
\centering
\caption{List of Symbols}
\begin{tabular}{|c|c|}
\hline
$T_d$  & Number of discrete time periods \\
\hline
$P_{sp}(t_d)$ & Spectrum availability probability of BSs at period $t_d$ \\
\hline
$f_{cp}(l,t_d)$ & Probability density function of CPU cycle at period $t_d$  \\
\hline
$\mathcal{J}$ & eVTOL Trajectory\\
\hline
$N(\mathcal{J})$ & Number of selected BSs (BS in $\mathcal{J}$) \\ 
\hline
$M(\mathcal{J})$ & Number of travel intervals in $\mathcal{J}$ \\ \hline
$P_h$ & Hovering-related energy consumption \\ \hline
$P_f(v)$ & Propulsion-related energy consumption \\ \hline
$T_h(n)$ & eVTOL's hover time for BS $n$ \\ \hline
$T_f(m)$ & eVTOL's flight time from $m$-th to $m+1$-th BS \\ \hline
$T_{com}$ & Task transmission time \\ \hline
$T_{cal}$ & Task processing time \\ \hline
\end{tabular}
\label{Used_Symbols}
\end{table}


While existing studies have advanced UAV/eVTOL communication and trajectory planning methodologies, few address the synchronized optimization of task offloading and flight trajectory under dynamic BS resource constraints, including time-varying spectrum and computational availability. Moreover, the interplay between BS distribution, dynamic BS resources (spectrum and CPU cycles), and eVTOL task latency tolerance presents significant analytical and optimization challenges.
To bridge this critical gap, this work propose an MCTS-based method that unifies temporal CR spectrum access strategies with spatial trajectory optimization problem. This co-design achieves simultaneous maximization of spectrum access success probability and task completion efficiency, addressing both temporal and spatial resource constraints.
Additionally, Tab.~\ref{Used_Symbols} summarizes the key notations and symbols used throughout this paper.
}

\section{System Model}
\begin{figure} [h]
\centering
\includegraphics[width=0.48\textwidth]{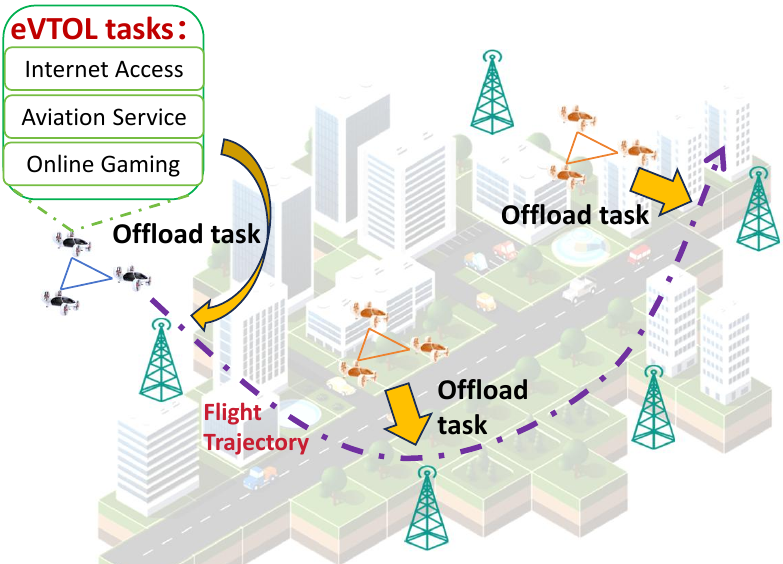} 
\caption{Illustration of the eVTOL task offloading scenario.}
\label{Scenario_picture}
\end{figure}

We consider a scenario where the eVTOL follows a predetermined trajectory, during which passengers onboard request internet access and entertainment services, as shown in Fig.~\ref{Scenario_picture}. 
To support these demands, the eVTOL must communicate with nearby BSs for computational task offloading and connectivity. 
However, due to limited encounter durations or constrained BS resources (e.g., spectrum availability and CPU cycles), a single BS may be insufficient to complete the offloading requirements. Consequently, the eVTOL needs to distribute its computational tasks across multiple BSs along its trajectory. It is worth noting that this work assumes task continuity across BSs, i.e., uncompleted tasks are resumed at subsequent BSs. Inter-BS cooperation is considered beyond the scope of this study, as the primary focus lies in developing an algorithm that addresses BS resource estimation and trajectory optimization.

\begin{figure} [h]
\centering
\includegraphics[width=0.48\textwidth]{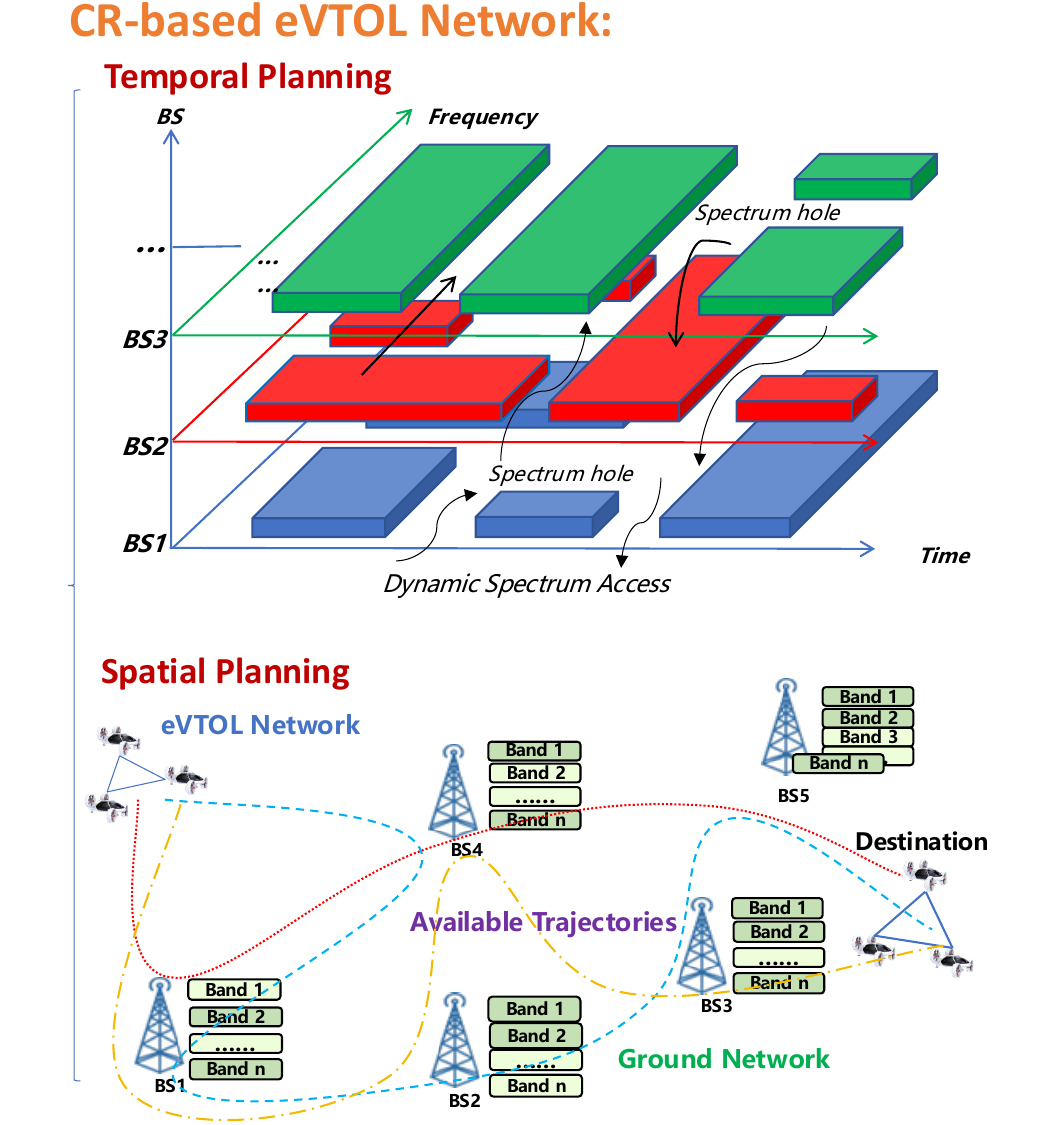} 
\caption{Illustration of the CR-based eVTOL network.}
\label{CR_AAM_scenario}
\end{figure}

The system considered in this work comprises $N$ BSs, several ground users (primary users), and an eVTOL swarm acting as the secondary user. Leveraging the spectrum-sharing and resource management principles of CR networks, the swarm optimizes its trajectory to avoid resource conflicts with primary users while minimizing its total energy consumption. A BS is considered available if it has idle spectrum (referred to as a spectrum hole, as illustrated in Fig. \ref{CR_AAM_scenario}) and sufficient CPU cycles to assist the swarm with computational tasks.
BS availability fluctuates over time due to varying demand from primary users; for example, fewer resources are typically available during peak hours (e.g., commuting times) compared to late-night periods. To model these temporal dynamics, we divide the day into $T_d$ discrete time period and denote the current time period by $t_d \in \{1, \cdots, T_d\}$. Let $\mathbf{P}_{sp}(t_d) = [P_{sp,1}(t_d), \cdots, P_{sp,N}(t_d)]$ denote the vector of spectrum availability probabilities across all BSs at time period $t_d$, and let $f_{\rm cp}(l,t_d)$ represent the probability density function (PDF) of available CPU cycle durations at the same time.
In this work, we assume that the elements of $\mathbf{P}_{sp}(t_d)$ follow Gaussian distributions with different means and variances \cite{bsgd}, i.e., $P_{sp,i}(t_d) \sim \mathcal{N}(\mu_i(t_d), \sigma^2_i(t_d))$ for $i \in \{1, \cdots, N\}$, where $\mu_i(t_d)$ and $\sigma^2_i(t_d)$ denote the mean and variance, respectively.

As the swarm travels toward its destination, e.g., from $S_b$ to $S_e$, it selects a subset of available BSs for access and task offloading. 
Note that the swarm does not need to visit all $N$ BSs, instead, it chooses a subset based on a joint consideration of spectrum availability probabilities, the duration of idle CPU cycles for processing tasks, and the energy cost associated with visiting each BS (further details are provided in the Subsection \ref{subsec_traj}).
Let $\mathcal{J}$ denote the trajectory taken by the swarm, during which it visits $N(\mathcal{J})$ BSs and traverses $M(\mathcal{J})$ travel intervals. Each travel interval refers to the movement from BS $i$ to BS $j$, where $i, j \in \mathcal{J}$. During wireless access and task offloading, the swarm is assumed to hover above the selected BSs. The overall network topology is illustrated in Fig. \ref{CR_AAM_scenario}.

\subsection{Energy Consumption}
Consider the onboard battery of the eVTOLs as one of the primary limitations of the system. eVTOLs rely on their internal batteries for power supply, and as a result, their flight or hovering time is constrained. 
The energy consumption model in this work is based on \cite{pcm}. Let $P_h$ denote the hovering-related energy consumption, which includes both the hovering and communication-related power, and let $P_f(v)$ represent the propulsion-related energy consumption, where $v$ is the velocity of the swarm. For simplicity, we use a single eVTOL to represent the energy of all eVTOLs in the swarm.

Let $T_h(i) = T_{\rm com}(i) + T_{\rm cal}(i)$ represent the time the swarm hovers above BS $i$, where $T_{\rm com}(i)$ is the task offloading time and $T_{\rm cal}(i)$ is the task processing time. The flight time from BS $i$ to BS $j$, denoted as a travel interval $M_{i}(\mathcal{J}) \in M(\mathcal{J})$, is represented as $T_f(i)$. Furthermore, we define the total traveling time $T$, which represents the overall time the swarm spends along its entire route. Thus, the total traveling and hovering time must satisfy the following constraint,
\begin{equation}
\begin{aligned}
\sum_{i}^{N(\mathcal{J})} T_h(i) +\sum_{i}^{M(\mathcal{J})} T_f(i) \leq T.
\label{vsdgjeguiener}
\end{aligned}    
\end{equation}
In addition, given the limited onboard battery, the energy consumed for hovering, communication, and propulsion must be strictly less than the battery capacity,
\begin{equation}
\begin{aligned}
\mathcal{W}_t = \sum_{i}^{N(\mathcal{J})} P_h T_h(i) + \sum_{i}^{M(\mathcal{J})} P_f(v) T_f(i) \leq B_{\rm max},
\label{eq_energyconstrains}
\end{aligned}    
\end{equation}
where $B_{\rm max}$ denotes the battery capacity.

\subsection{BS Temporal Resource Estimation}

The objective of this work is designed to generate an energy-efficient eVTOL trajectory that maximizes the data offloading success probability, subject to energy and time constraints.

Specifically, the trajectory design process in this work consists of two phases: (i) the temporal phase and (ii) the spatial phase.
The first phase estimates the BS resource availability at different time period during the travel duration. 
The second phase designs the swarm's trajectory to minimize energy consumption during hovering and traveling, while maximizing the probability of successful access to BSs (e.g., ensuring the swarm reaches a BS with available spectrum).

\begin{remark}
\label{rem_st_dif}
Note that these two phases interact with each other. As the swarm travels toward the target BS, the availability of BSs changes over time. Additionally, upon the swarm's arrival at the target BS, both the distances to surrounding BSs and the availability of those BSs may change. Therefore, if the swarm fails to offload the task to the planned BS, a new trajectory plan must be developed, taking into account the updated temporal BS availability and spatial locations.
\end{remark}


Recall that spectrum availability varies with time periods and BSs. Therefore, we propose a Multi-Armed Bandit (MAB) model to pre-learn the temporal resource distribution pattern of different BSs.
\begin{definition}[BS Temporal Resource Estimation] 
\label{def_resource_est}
The eVTOL swarm pre-learns the BS's temporal resource availability using a MAB model. Specifically, at time period $t_d$, the estimated probability of access to BSs is denoted as $\mathbf{\hat{P}}_{sp}(t_d)$.
\end{definition}
\begin{remark}
Note that, in Definition \ref{def_resource_est}, we only consider the eVTOL swarm's ability to pre-learn the probability of successful access to BSs. This is because, due to the limited information available to the swarm, the mean available CPU cycle $\mathbf{\hat{L}}_{cp}$ of a BS can only be obtained when the swarm is connected to the BS. Therefore, it is infeasible to obtain this information during the pre-learning stage.

The fundamental problem addressed by the MAB scenario is as follows: There are $N$ machines (BSs) in the environment, and pulling the lever of a machine yields a reward. 
A gambler (eVTOL swarm) faces these $N$ machines but does not initially know the actual reward distribution of each machine. 
The gambler must maximize total gains throughout the game by making choices based on past selections and feedback (rewards), which fits well with the scenario considered in this work.

\end{remark}

\subsection{Spatial-temporal Trajectory Optimization}
\label{subsec_traj}
{\color{black}
Let $\mathcal{A}$ denote the access selection of the eVTOL swarm, which includes sequence decisions $a_i(n)\in\mathcal{A}$, where $a_i(n)=1$ for $i \in \{1, \cdots, N\}$ and $n \in \{1, \cdots, N(\mathcal{J})\}$, corresponding to select $i$-th BS to be visited at $n$-th selected stage. Otherwise, $a_i(n) = 0$. The specific $\mathcal{A}$ in each selected stage is controlled by $\mathbf{\hat{P}}_{sp}(t_d)$, the estimated access probability of eVTOL swarm for each BS. Consequently, the outcome trajectory is defined as $\mathcal{J} = \{ j_1,j_2,\dots, j_{N(\mathcal{J})}  | a_{j_n}(n)=1,  n\in\{1,\dots, N(\mathcal{J})\} \}$. Furthermore, the trajectory optimization problem is formulated as,


\begin{equation}
\begin{aligned}
\mathcal{P}1: &\min_{\mathbf{T_h, T_f},\mathcal{A}} \ \ \sum_{n=1}^{N(\mathcal{J})}P_{h}T_{h}(n) + \sum_{m=1}^{M(\mathcal{J})}P_{f}(v)T_{f}(m) - \lambda \eta, \\
\text{s.t.} \ \ 
&C_1:\sum_{n=1}^{N(\mathcal{J})}T_{h}(n) + \sum_{m=1}^{M(\mathcal{J})}T_{f}(m)\leq T, \\
&C_2:T_{h}(n) = T_{\rm com}(n) + T_{\rm cal}(n),n \in [1,N(\mathcal{J})], \\
&C_3:\eta \in [0,1], \\
&C_4:\|q[t]-q[t-1]\|\leq V_{\mathrm{max}}, \forall t \in [0,T], \\
&C_5:\|v[t]-v[t-1]\|\leq a_{\mathrm{max}}, \forall t \in [0,T], \\
&C_6:q[0] = S_{b}, ~q[T] = S_{e}, \\
&C_7:\mathcal{W}_t \leq B_{\rm max},\nonumber
\end{aligned}
\label{p1}
\end{equation}
\noindent where $\lambda$ is the weight coefficient, and $\eta$ is the successful offloading probability (offloading completion probability). In addition, the variables $q[t]$ and $v[t]$ represent the location and velocity of the eVTOL swarm at time $t$, respectively. $a_{\rm max}$ denotes the maximum acceleration of the eVTOL swarm.

The optimization objective is designed to minimize the difference between the weighted sum of energy consumption and the successful offloading probability, namely, to minimize energy consumption while maximizing task completion probability.
The decision variables of this optimization problem include the hover time $\mathbf{T_h} = \{T_h(1),\dots, T_h(N(\mathcal{J}))\}$ for task offloading to BSs (the total number of the selected BSs is $N(\mathcal{J})$), the flight time $\mathbf{T_f} = \{T_f(1),\dots, T_f(M(\mathcal{J}))\}$ for each the intervals between adjacent BSs in the trajectory $\mathcal{J}$ (the number of the path segments is $M(\mathcal{J})$), and the action set $\mathcal{A}$. Note that the hover time $\mathbf{T_h}$ relies on the access probability and completion time of task offloading. 
The flight time $\mathbf{T_f}$ is determined by the flight velocity of the eVTOL swarm.

Here, $C_1$ represents that the total time requirement. The eVTOL travel comprises two components: the flight time between BSs and the hovering time at BSs. The flight time refers to the cumulative duration of movement between the BSs selected in the action set \(\mathcal{J}\), while the hovering time is the cumulative duration of staying at each selected BS. To ensure the eVTOL task is completed within the specified time limit, the total time must not exceed the maximum allowable mission duration $T$. 

While $C_2$ captures the hovering time. It consists of the offload transmission phase (taking \(T_{\text{com}}\), the time for the eVTOL to transmit task data to the BS) and the BS processing phase (taking \(T_{\text{cal}}\), the time for the BS to receive and compute the data). 
The hovering time must fully cover these two phases. 

$C_3$ reflects the probability characteristic of task offloading success rate. 
The outcome of task offloading has only two discrete states: "success" or "failure". When successful, the task requirements are fully met (\(\eta = 1\)); when failed, the task remains incomplete (\(\eta = 0\)), with no intermediate states. Consequently, the task success rate \(\eta\) is limited to a value range of $0$ to $1$. 

$C_4$ represents the eVTOL velocity limitation. The norm of the position difference between adjacent time steps must not exceed a maximum safe velocity \(V_{\text{max}}\).
$C_5$ is the eVTOL acceleration limitation. Based on the eVTOL’s flight stability requirements, the norm of the velocity difference between adjacent time steps must not exceed a maximum safe acceleration \(a_{\text{max}}\).
$C_6$ is the initial and destination Location Constraint. The eVTOL’s mission has clear spatial boundaries: the mission starts from a specified initial location \(S_{b}\) and ends at a specified destination \(S_{e}\).
$C_7$ is the total Energy Consumption Constraint. Since the eVTOL relies on an onboard battery for power, the total energy consumption must not exceed the maximum battery capacity \(B_{\text{max}}\).

}

{\color{black} The joint trajectory and CR-based task offloading problem ($\mathcal{P}1$) is formulated as a Mixed-Integer Nonlinear Programming (MINLP) problem. Its complexity mainly stems from three aspects. Firstly, mixed variable types:
There are \textit{Binary Selection Indicators:} $a_i(n) \in \{0,1\}$ for $i \in \{1, \cdots, N\}$ and $n \in \{1, \cdots, N(\mathcal{J})\}$, where $a_i(n) = 1$ denotes that the eVTOL swarm visits $i$-th BS at the $n$-th selected stage, and $a_i = 0$ otherwise. These variables determine the BS access sequence $\mathcal{J}$.
Additionally, \textit{Continuous variables:} The 3D position $q[t] \in \mathbb{R}^3$, velocity $v[t] \in \mathbb{R}^3$, hover time $T_h$, and flight time $T_f$ are continuous-valued and define the spatio-temporal trajectory.
This hybrid structure inherently increases the complexity of the solution space. 

Secondly, a nonlinear objective function: the objective function integrates energy consumption and successful offloading probability in a nonlinear manner. 
Specifically, the energy term involves a quadratic function of velocity (via propulsion power \(P_f(v)\)), while the successful offloading probability \(\eta\) is nonlinearly dependent on spectrum availability and CPU cycle dynamics. This coupling makes convex reformulation of the problem difficult and further exacerbates its nonlinearity.

Finally, polynomial growth of constraints: the number of constraints scales polynomially with both the number of candidate BSs $N$ and the number of time period $T$. 
For example, the time budget (C1), kinematic constraints (C4–C5), and energy budget (C7) must be satisfied at each time step, resulting in a constraint evaluation volume of \(O(NT)\). Therefore, the growth scale of its search space is approximately \(O(N^\kappa)\) (where $\kappa$ represents the effective depth of the decision tree, such as the number of time slots or sequential decisions), making exhaustive search completely infeasible in deployment.

In what follows, we begin the optimization analysis. The goal of this work is to obtain the optimal eVTOL trajectory that minimizes energy consumption while maximizing successful task offloading. 
To achieve this, we first utilize a MAB model to estimate the temporal spectrum availability of the BSs (defined in Definition \ref{def_resource_est} and solved in Subsection \ref{subsec_prediction}, Alg.\ref{Ag1}). Based on the MAB estimated information, we propose a Monte Carlo Tree Search (MCTS)-based spatial-temporal decision-making algorithm (Subsection \ref{subsec_MCTS}, Alg.\ref{Ag2}) to design the eVTOL's trajectory and task offloading operations, ultimately solving $\mathcal{P}1$.
}


\section{Performance Analysis}
\begin{figure}[h]
\centering
\includegraphics[width=0.48\textwidth]{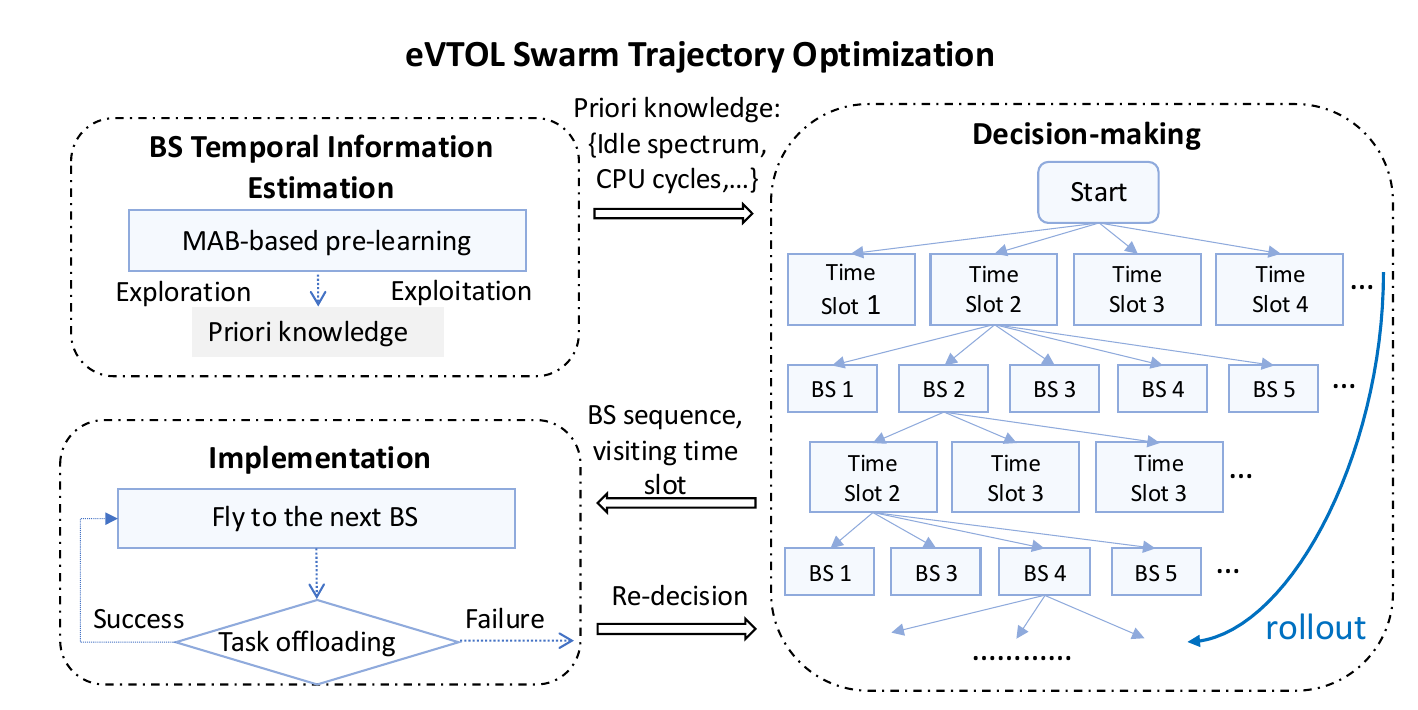} 
\caption{Illustration of the trajectory optimization model.}
\label{fig_2}
\end{figure}

To address the spatio-temporal challenges discussed in Remark \ref{rem_st_dif}, we propose a three-stage trajectory planning algorithm, as illustrated in Fig.~\ref{fig_2}: (i) Stage 1: BS temporal resource availability estimation (or pre-learning), (ii) Stage 2: spatial-temporal decision-making, and (iii) Stage 3: implementation.

The goal of the BS temporal resource availability estimation stage is to use the MAB model to pre-learn (or estimate) the spectrum availability at each BS over different periods of the day. Based on this pre-learned information, Stage 2 employs an MCTS-based algorithm to optimize the swarm’s trajectory, considering spectrum availability, CPU cycles, remaining tasks, energy, and time constraints. In the implementation stage, the swarm follows the planned trajectory and offloads tasks at the BSs. After each MCTS iteration, the trajectory is updated based on the task offloading status, current locations, and updated BS resource information. This process continues until all tasks are completed or until the time limit $T$ and energy capacity $B_{\rm max}$ are exhausted. Further details are provided in the following sections.

\subsection{BS Temporal Resource Information Estimation}
\label{subsec_prediction}
Recall that the primary objective in this stage is to accurately estimate the spectrum availability probability $\mathbf{P}_{sp}(t_d)$ at each BS over different time periods $t_d$. Since eVTOLs lack prior knowledge of these parameters, they must gather spectrum availability information through a pre-learning stage, as well as CPU cycle durations during actual connections with each BS. To facilitate more accurate decision-making in subsequent stages, eVTOLs must continuously explore and evaluate multiple BSs over time.

Generally, the MAB framework models the trade-off between {\em exploration} and {\em exploitation} when selecting from several uncertain options. Mathematically, at each time step, an agent (eVTOL network) selects an arm (BS) and receives a reward (spectrum availability) drawn from an unknown distribution. The goal is to maximize the cumulative expected reward over time.
\begin{lemma}[MAB Model]
Recall that $\mathcal{A}$ denotes the action set representing associations with BSs, and the reward for selecting an action $a \in \mathcal{A}$ is given by $r(a)$. Specifically, $r(a) = 1$ if the connection is successful, and $r(a) = 0$ otherwise. The probability of $r(a) = 1$ is unknown and varies with BSs and time of a day, and let $R(r|a)$ be the distribution of $r(a)$. After pre-learning over multiple iterations, the frequency of $r(a) = 1$ converges to $P_{sp,a}(t_d)$, which is denoted by $\hat{P}_{sp,a}(t_d)$.

Moreover, exploitation refers to selecting the BS with the currently highest estimated reward, while exploration involves choosing BSs that have been visited less frequently in previous rounds to improve future estimates. Typically, only one BS is selected at each time step.
\end{lemma}


The goal of the MAB model is to maximize the cumulative reward over the travel duration $T$, which is,
\begin{equation}
\begin{aligned}
{r} = \sum_{t=1}^T r_t, \ r_t \sim R(\cdot |a_t),
\end{aligned} 
\label{sadheygafew}
\end{equation}
where, $a_t$ denotes the action of selecting a specific BS at time $t$, and $r_t$ represents the reward obtained from action $a_t$. For each action $a$, we define its expected reward as,
\begin{align}
\theta(a_t) &= \mathbb{E}_{r \sim R(\cdot |a_t)}[r]\nonumber\\
&=\frac{1}{N_{t}(a_t)} \sum_{\tau=1}^{T} r_{\tau} \mathbb{I}_{a_t}(a_{\tau}),
\label{fxcgkldrfjgdrlk}
\end{align}
where $\mathbb{E}[\cdot]$ is the expectation operation, $\mathbb{I}$ is the indicator function, $\mathbb{I}_a(b) = 1$, iff $b = a$, and  $N_{t}(a)$ represents the total times that action $a$ has been taken up to time $t$. 

\begin{lemma}[Expected Reward and Regrets of the MAB Model]\label{lem_rewardandregret}
The optimal expected reward is expressed as 
\begin{align}
\theta^* = \max_{a_t\in \mathcal{A}}\theta(a_t),
\end{align}
and the regret of the MAB model, which is defined as the difference in expected reward between the current BS selection and the optimal BS selection, given by $\theta^*-\theta(a_t)$, and the cumulative regret is the total accumulated regret after $T$ times BS selections. Let ${a_1, a_2,\dots, a_{T_{st}}}$ be the completed $T_{st}$-step decision sequence, the expected cumulative regret is given by,
\begin{equation}
\begin{aligned}
    \mathcal{L}_{T_{st}} = \mathbb{E}\left[\sum_{t=1}^{T_{st}} [\theta^*-\theta(a_t)]\right].
\end{aligned} 
\label{gfhfghdrt}
\end{equation}
\end{lemma}



To solve the optimization problem in Lemma \ref{lem_rewardandregret}, we adopt the Upper Confidence Bound (UCB) method to estimate the spectrum availability probability for each BS, denoted by $\hat{\mathbf{P}}_{sp}(t_d)$, based on historical operation data. The UCB approach balances exploration and exploitation by prioritizing actions with higher uncertainty, thereby encouraging the selection of less-frequently visited BSs to improve future decision-making.

\begin{lemma}[UCB Solution to the MAB Model]
\label{lem_UCB_sol}
Let  $U_{t}(a)$ be the uncertainty of $a$. The goal of UCB method is to  maximize the action confidence which is given in \cite{9397771Wang} as
\begin{align}
    a_{t}^{\rm UCB}=\arg \max _{a \in \mathcal{A}} \theta(a_t)+\eta_c U_{t}(a),
\label{dkjfhbrfbkuas}
\end{align}
where $\eta_c$ is the scaling coefficient, and
\begin{align}
U_{t}(a) = \sqrt{\frac{2 \log t}{N_{t}(a_{t})}}.
\end{align}
\end{lemma}
\begin{remark}[Brief Description of the UCB solution]
At the beginning of action selection $a$ (i.e., choosing a BS), the initial confidence value $a_{t}^{\rm UCB}$ is set to zero. The eVTOL swarm randomly selects one action and receives a corresponding reward.

On the second trial, the previously chosen action has lower uncertainty and thus a smaller $U_{t}(a)$, making it less likely to be selected again. Instead, the agent (eVTOL swarm) selects a different action $a$ from the remaining set $\mathcal{A}$, continuing this process until each $a_i \in \mathcal{A}, i \in {1,\cdots,N}$ has been tried at least once. This initial phase is commonly referred to as the exploration phase, enabling the agent to develop a global understanding of the environment.

In subsequent iterations, since all BSs have been selected at least once, the uncertainties across actions are initially equal, i.e., $U_{t}(a_i) = U_{t}(a_j), \forall i,j \in {1,\cdots,N}$. From this point of view, the BS selection becomes increasingly influenced by the estimated reward $\theta(a_{t})$, favoring exploitation. However, as certain BSs are selected more frequently, the term $\sqrt{\frac{2 \log t}{N_{t}(a)}}$ decreases, making less-frequently chosen BSs relatively more attractive for future selection.

In summary, the UCB strategy follows a simple yet effective rule: the greater the uncertainty associated with an action, the larger its confidence bound. As the number of iterations increases, the confidence intervals for all actions gradually converge, thus maintaining a balance between exploration and exploitation.
\end{remark}

The detailed steps of the UCB-MAB algorithm is presented in Alg.~\ref{Ag1}.






\begin{algorithm} 
    \caption{MAB Experience-obtained Algorithm} 
    \label{Ag1}
    \textbf{Input}: Number of experiments, Reward function (\ref{fxcgkldrfjgdrlk}), Scaling coefficient $\eta_c$; 
    
    \textbf{Output}: Each station's idle probability that's being calculated with different algorithms; the difference between algorithms and best solution
    
    \textbf{Initialization}: eVTOL's learning results; number of eVTOL's visiting\\
    Formulate a tree structure\\
    \For{$index$ = 0,1,2,...,N-1} {
	\For{$eposide$ =0,1,2,...,M-1} {
		Generate a random number\\
        Compare the random number with the parameters of greedy algorithm\\
		\If {random number $\leq$ $\epsilon$} {
                Explore and update
                }
        \Else {
                Exploit and update
                }
        Calculate each station's UCB value\\
        Choose the best BS\\
        Exploit and update\\
        }
    }
    Compare different algorithm's learning results and reward function\\
\end{algorithm}


\subsection{Spatial-temporal Trajectory Designing}
\label{subsec_MCTS}

In previous subsection, a MAB model is used to obtain BS temporal resource information $\hat{\mathbf{P}}_{sp}(t_d)$, and in this subsection, we combine the estimated temporal information with the spatial location of the eVTOL swarm, and propose a MCTS-based spatial-temporal trajectory design algorithm, and it can be summarized as follow four steps.
\begin{itemize}
\item Selection: Starting from the root node, the algorithm recursively selects the optimal child node until reaching a leaf node.
\item Expansion: If the leaf node is not a terminal node (i.e., the game has not ended), one or more child nodes are created, and a node is selected for further exploration.
\item Simulation: A simulated rollout is executed from the selected node until the game reaches a terminal.
\item Backpropagation: The results of the simulation are propagated back through the tree to update the counts and immediate reward of the current node sequence.
\end{itemize}


The first step of the spatio-temporal decision-making process is to select an appropriate time period for task offloading, taking into account factors such as the required offloading duration and the current spectrum availability at each BS.
Once the task offloading time period is determined, the second step involves selecting a set of BSs that can support the eVTOL swarm in completing its tasks. This decision is based on the estimated spectrum availability $\hat{\mathbf{P}}_{sp}(t_d)$ obtained in Algorithm~\ref{Ag1}, along with the BSs’ spatial locations, and the swarm’s energy and time constraints.

When selecting a time period, the previously collected data is used to identify the time period with the highest probability of having available spectrum,
\begin{align}
t_{d}^{\star} =\argmax_{t_d} ~\mathbf{\hat{P}}_{sp}(t_d).
\end{align}


As previously mentioned, the decision-making stage goes beyond considering only the estimated spectrum availability $\hat{\mathbf{P}}{sp}(t_d)$. It must also account for the eVTOL swarm’s energy and time constraints, as well as their current locations. To incorporate these additional factors, we modify the  expected reward in (\ref{fxcgkldrfjgdrlk}) within the MCTS framework by including the estimated CPU processing time $\hat{\mathbf{L}}{cp}(t_d)$ along with the swarm's energy and time limitations.
\begin{lemma}[Modified UCB Expected Reward]
\label{le_modifiedUCBreward}
Consider a specific travel interval $i \in M(\mathcal{J})$, represented by $M_i(\mathcal{J})$, where $M(\mathcal{J})$ denotes the set of travel intervals between BSs.
Let $\theta^{\prime}(a)$ be the modified expected reward, which characterized by weighted sum of three key factors: the estimated spectrum availability $\hat{\mathbf{P}}_{sp}(t_d)$, the estimated CPU processing time $\hat{\mathbf{L}}_{cp}(t_d)$, and the energy consumption of the eVTOL swarm associated with the travel interval,
\begin{align}
    \theta^{\prime}(a)=c_{1}\mathbf{\hat{P}}_{sp}(t_{d}^{\star}) + c_{2}\mathbf{\hat{L}}_{cp}(t_{d}^{\star}) + c_{3}\mathcal{W}_i ,
\label{rtweguhwreugh}
\end{align}
where $c_{1}$, $c_{2}$, and $c_{3}$ are weighting coefficients reflecting the relative priority of the three reference factors. For instance, in certain scenarios, energy consumption may take precedence over processing time.
The term $\mathcal{W}_i$ represents the total energy consumption associated with travel interval $M_i(\mathcal{J})$, which includes propulsion-related energy $P_h T_h(i)$ and the energy consumed during hovering and communication, given by $P_f(v) T_f(i)$, namely,
\begin{align}
    \mathcal{W}_i &= P_h T_h(i)+P_f(v)T_f(i)\nonumber\\
   & \stackrel{(a)}{=}\frac{|M_i(\mathcal{J})|}{\bar{v}}P_f(\bar{v}) + \mathbf{\hat{L}}_{cp}(t_{d}^{\star})P_h,
\end{align}
where step $(a)$ follows from the assumption that the hovering time $T_h(i)$ corresponds to the available CPU cycle duration—i.e., the eVTOL swarm and the BS complete task offloading, data processing, and result transmission within the BS's available CPU cycles.
The flight time is given by $T_f(i) = \frac{|M_i(\mathcal{J})|}{\bar{v}}$, where $|M_i(\mathcal{J})|$ denotes the distance between BS $i$ and BS $j$, and $\bar{v}$ is the average velocity.
Note that, due to the energy constraint of the eVTOL swarm, the following condition must be satisfied,
\begin{align}
\sum_{i}^{N(\mathcal{J})} \mathcal{W}_i = \mathcal{W}_t \leq B_{\rm max}.
\end{align}
\end{lemma}

While the eVTOL swarm uses the modified UCB expected reward function defined in Lemma~\ref{le_modifiedUCBreward} (\ref{rtweguhwreugh}) and iteratively applies the UCB solution to the MAB model as described in Lemma\ref{lem_UCB_sol} (\ref{dkjfhbrfbkuas}), it can estimate the probability of spectrum availability and the mean CPU cycle duration at each BS. However, upon arrival at a selected BS, the swarm may not always obtain the expected spectrum access or sufficient CPU cycles. In such cases, when the BS fails to deliver the expected resources, the eVTOL swarm must re-assess its strategy using the updated resource status, as well as its current energy and time constraints, and repeat the decision-making process accordingly.

Upon successfully completing each task, the eVTOL collects and updates statistical information on available spectrum and CPU cycle length based on their tree search experience. 
This allows them to continuously refine their understanding of the spectrum and CPU availability across different time periods and BSs.

Through multiple explorations and refinements, the tree structure gradually expands and improves, ultimately forming a comprehensive and accurate decision tree. Using this decision tree, the eVTOL swarm can make optimal decisions following the same process.

\begin{figure} [h]
\centering
\includegraphics[width=1\columnwidth]{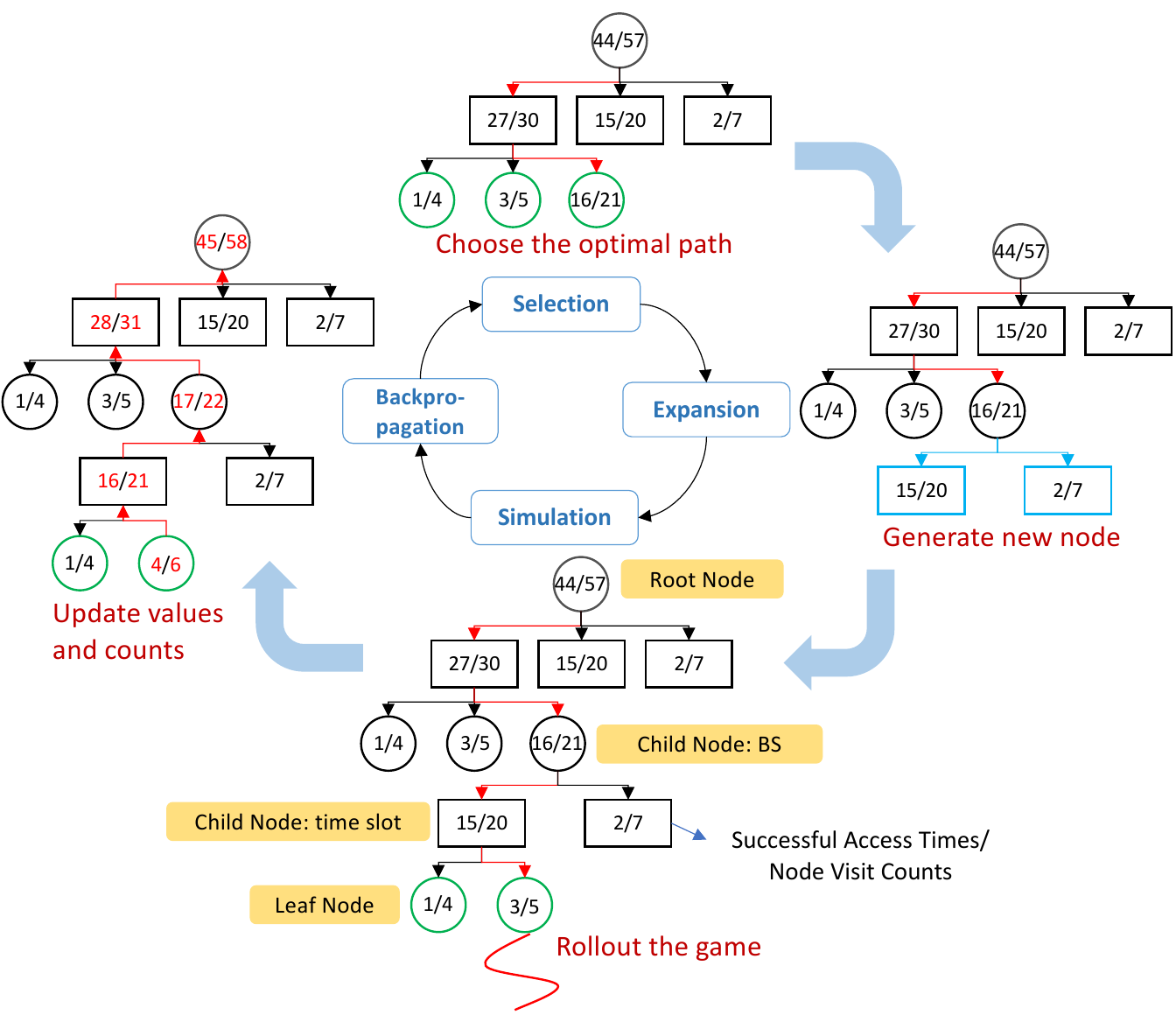} 
     \caption{Illustration of the MCTS-based model diagram.} 
\label{fig_4}
\end{figure}

\begin{lemma}[MCTS-based Model] Let the root node represent the eVTOL swarm, and the leaf nodes denote the final BSs in the visiting sequence where task offloading is successfully completed. The intermediate nodes represent either time periods or BSs, as eVTOLs must first select a time period before choosing a BS. However, since not all BSs may support successful task offloading, the decision process may need to continue, dynamically expanding the subtree to explore additional time period and BS combinations until a successful offloading opportunity is found. The MCTS-based decision-making trajectory planning method is provided in Algorithm \ref{Ag2} and an example is provided in Example \ref{exp1}, as well as illustrated in Fig. \ref{fig_4}.
\end{lemma}
\begin{remark}
In this process, each node records two key pieces of information: the estimated immediate reward derived from rollout simulation results and the total number of times the node has been visited. To ensure a more efficient and memory-saving implementation of the algorithm, only one leaf node is added in each expansion.
\end{remark}

\begin{example}
\label{exp1}
Fig.~\ref{fig_4} illustrates an example of the MCTS-based model. At the beginning of the selection process, the modified MAB model has been trial 57 times, out of which 44 attempts were successful. This is reflected at the top node of the figure, where the fraction $44/57$ is shown inside the circle. Beneath this root node, three square nodes represent three different time periods (for simplicity, only three are shown in this example), with corresponding success ratios of $27/30$, $15/20$, and $2/7$, respectively. These ratios indicate the empirically estimated success probabilities for accessing the spectrum during those specific time periods. Clearly, the first time period offers the highest access probability, so the eVTOL swarm prefers this time period for task offloading, as indicated by the red arrow. Below the square node, three additional circles represent the three BSs (again, only three are shown for illustration) that have not yet been connected. However, based on the pre-learning phase, the swarm has gathered prior knowledge of these BSs, with estimated success probabilities of $\hat{P}_{sp} = \{1/4,3/5,16/21\}$,  respectively. As a result, the swarm chooses the third BS, which has the highest success access probability.

Once the swarm completes its first selection, implements the decision, and flies to the selected BS, the MCTS-based tree expands. Suppose the time period changes, and the swarm has only two options left, i.e., time periods with success access probabilities of $15/20$ and $2/7$. The swarm follows the same process and chooses the BS that provides the highest $\hat{P}_{sp} = 3/5$.  If the task offloading is successfully completed, i.e., the BS provides enough CPU cycles, the process ends. Otherwise, the swarm updates the collected information, such as total trials and successful attempts, and the process continues. 
\end{example}

{\color{black}
Hereafter, we present the phase-wise time complexity analysis for the proposed MCTS algorithm. 
During the expansion phase, adding child nodes to the search tree incurs constant time complexity $O(1)$. The selection and backpropagation phases exhibit similar complexity characteristics, both scaling with the tree's branching factor $b$ and total node count $t$, resulting in \(O(b \log_b t)\) complexity for each phase. This logarithmic relationship reflects the tree's hierarchical structure, where the depth grows logarithmically with the number of nodes.

 The simulation phase, which simulates BS selection and task offloading (encompassing trajectory calculation), operates with polynomial complexity \(O(k)\), where $k$ represents either the number of simulation steps. Consequently, the overall time complexity across $n$ iterations is \(O\left( n \cdot \left( b \log_b t + k \right) \right)\).

 In practical eVTOL task offloading scenarios, this theoretical complexity remains computationally tractable for several key reasons. First, the number of candidate BSs and the constrained depth of trajectory decision trees naturally limit both $b$ and $t$ to manageable values. Second, the parameter $k$ in the simulation phase demonstrates polynomial rather than exponential growth, maintaining reasonable computational demands. Most significantly, the integration of MAB-based prior knowledge fundamentally enhances the algorithm's efficiency by guiding the selection process toward promising branches. This approach substantially reduces the effective search space during execution, ensuring the algorithm operates efficiently while maintaining high decision quality.
 }

\begin{algorithm} 
    \caption{MCTS-based Time Sequential Decision-making Algotirhm} \label{Ag2}
    \textbf{Input}: $\eta_c$, time limitation $T$
    
    \textbf{Output}: Flight trajectory $\mathcal{J}$, energy consumption $\mathcal{W}_t$, Task completion rate 
    
    \textbf{Initialization}: eVTOL's energy consumption, time spent \\
    Formulate tree structure \\
    \For{$phase$ =0,1,2,...$T_{p}$-1} 
    {
		Calculate each phase's UCB value\\
        Choose the best phase\\
        \For{$eposiode$ = 0,1,2,...M-1} {
        Calculate each station's UCB value\\
        Choose the best station\\
        \If{Task offloading failed} {
            Deselect the failed BS\\
            Update then go back to step 8\\
		}
        \If{Running out of time} {
            Deselect the time phase before\\
            Update then go back to step 5\\
        }
        \If{Finishing the task} {
            Fly to the next BS\\
        }
        }
    }				
\end{algorithm}

\section{Numerical Results}
In this section, we validate the proposed algorithm through simulations via MATLAB. Unless otherwise stated, the simulation parameters used in this work are provided in Table \ref{simulationargs}.
The simulation details are as follows.

 \begin{table}[!ht]
    \centering
    \caption{Simulation Parameters}
    \begin{tabular}{l|l|l}
    \hline
       \textbf{Symbol} & \textbf{Description} & \textbf{Value} \\ \hline
       $N$ & Number of BSs  &  $5$ - $10$ \\ \hline
       $/$ &Number of eVTOLs & $3$ - $5$  \\ \hline
       $V_{\rm max}$ & Maximum speed of eVTOL ($m/s$) & $30$ \\ \hline
       $a_{\rm max}$ & Maximum acceleration of eVTOL ($m/s^{2}$) & $20$ \\ \hline
       $\epsilon$ & Weight of $\epsilon$-Greedy & $0.5$ \\ \hline
       $\eta_c$ & Scaling coefficient of UCB & $1.0$  \\ \hline
       $T$ & Required total time for task processing (s) &  $40$ - $70$ \\ \hline
       $/$ & Required CPU cycle duration to complete tasks (s)  & $20$ - $40$ \\ \hline
    \end{tabular}
    \label{simulationargs}
\end{table}

In our simulation, the eVTOL aircraft obtains historical BS operation data from the Air Traffic Control (ATC) system prior to takeoff. Based on the collected data, the eVTOL swarm estimates the time-varying spectrum availability of BSs using the proposed MAB algorithm and determines the optimal flight time periods and BS access sequence for task offloading during flight using the proposed MCTS-based algorithm. We evaluate the algorithm’s efficiency under three different BS configurations: $5$, $7$, and $10$, along with varying CPU cycle durations. To ensure fair comparison across configurations, the required CPU cycle durations (i.e., computational time overhead) for task offloading are scaled proportionally to the number of BSs: 20 minutes (5 BSs), 30 minutes (7 BSs), and 40 minutes (10 BSs), respectively. This scaling reflects the increased spectrum and computing resources of BSs while maintaining proportional computational demands of eVTOL tasks.


Hereafter, we compare the proposed MCTS-based approach with two benchmark methods: i) {\em the classical Traveling Salesman Problem (TSP) algorithm}, which optimizes the eVTOL trajectory (e.g., solving $\mathcal{P}1$) by minimizing energy consumption and task failure probability through sequentially visiting all BSs via the shortest path; and ii) {\em the $\epsilon$-greedy algorithm}, which minimizes $\mathcal{P}1$ by selecting the optimal BS and time period for task offloading with probability $1 - \epsilon$, while exploring random BSs and time periods with probability $\epsilon$.

\begin{figure*}[ht]
\centering
\subfloat[]{\label{Access_Prob_5b}{\includegraphics[width=0.31\linewidth]{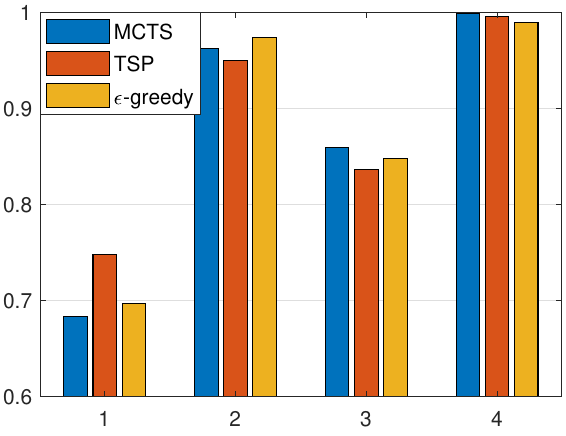}}} \hfill
\subfloat[]{\label{Access_Prob_7b}{\includegraphics[width=0.31\linewidth]{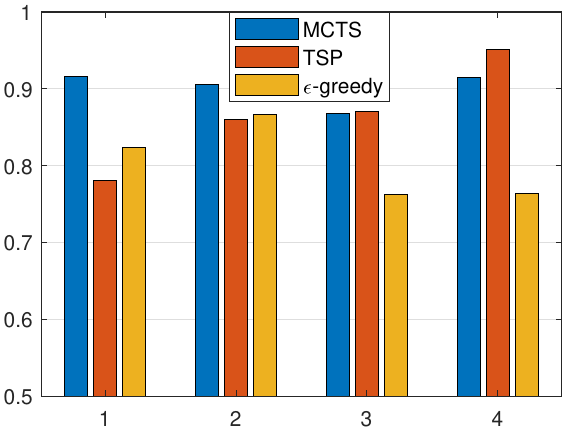}}} \hfill
\subfloat[]{\label{Access_Prob_10b}{\includegraphics[width=0.31\linewidth]{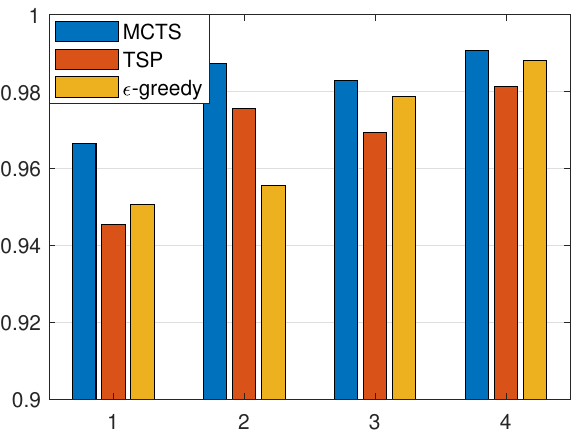}}} \hfill
\caption{Simulation results of the task completion probability, $\eta$, under varying numbers of BSs and CPU cycle durations, \textbf{(a)} $N = 5$ BSs with a required CPU cycle duration of 20 minutes, \textbf{(b)} $N = 7$ BSs with a required CPU cycle duration of 30 minutes, and \textbf{(c)} $N = 10$ BSs with a required CPU cycle duration of 40 minutes.}
\label{Access_Prob_integration}
\end{figure*}


\subsection{Task Completion Probability}

\begin{figure*}
\centering
\subfloat[]{\label{5BS_PC}{\includegraphics[width=0.31\linewidth]{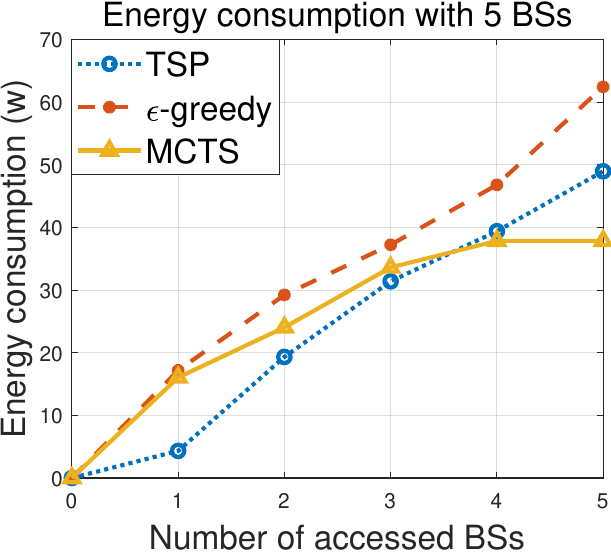}}} \hfill
\subfloat[]{\label{7BS_PC}{\includegraphics[width=0.31\linewidth]{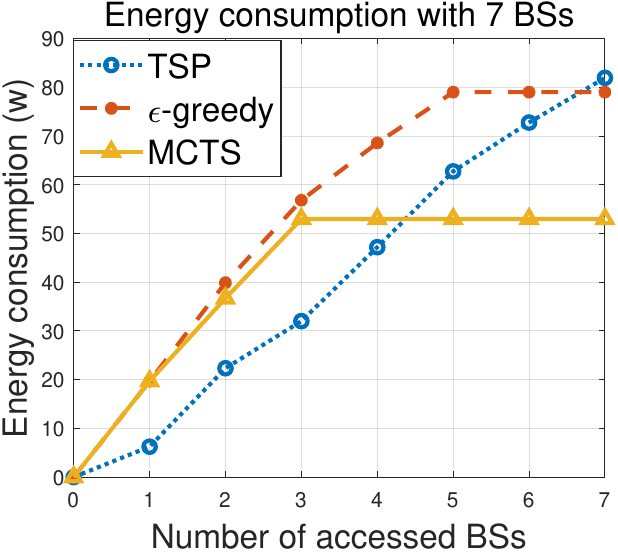}}} \hfill
\subfloat[]{\label{10BS_PC}{\includegraphics[width=0.31\linewidth]{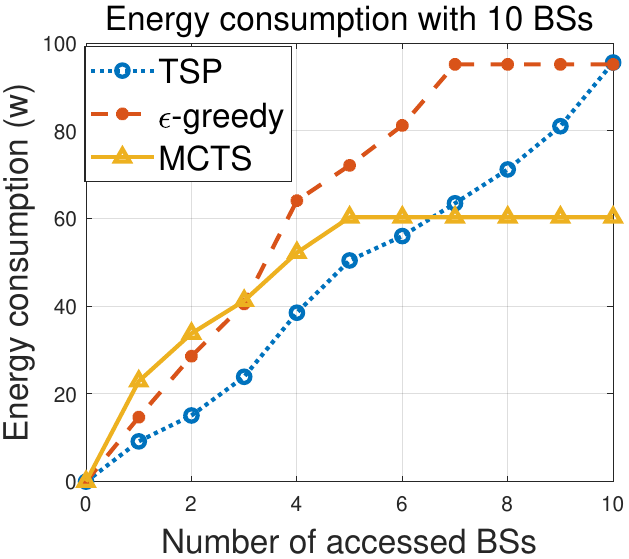}}} \hfill
\caption{Energy consumption comparison under different number of the accessed BSs while the network has \textbf{(a)} $N = 5$ BSs, \textbf{(b)} $N = 7$ BSs, \textbf{(c)} $N = 10$ BSs.}
\label{Power_Consum_integration}
\end{figure*}

In Fig.~\ref{Access_Prob_integration}, we plot the simulation results of $\eta$ under three different numbers of BSs and CPU cycle durations versus four different resource availability scenarios, respectively.

Fig.~\ref{Access_Prob_5b} illustrates $\eta$ when $N = 5$ BSs under four resource availability scenarios: \textbf{Scenario 1}: two BSs with $>80\%$ spectrum availability, two BSs providing $50\%$ required CPU cycles, \textbf{Scenario 2}: two BSs with $>80\%$ spectrum availability, five BSs providing $50\%$ CPU cycles, \textbf{Scenario 3}: five BSs with $>80\%$ spectrum availability, two BSs offering large CPU cycles, \textbf{Scenario 4}: five BSs with high spectrum availability ($>80\%$) and large CPU cycles.

The $x$-axis represents the four resource availability scenarios, while the $y$-axis shows the task completion probability $\eta$. The proposed MCTS-based algorithm demonstrates superior performance in resource-abundant environments (\textbf{Scenarios 3 and 4}), highlighting its ability to strategically prioritize BSs with sufficient resources rather than exhaustively visiting all BSs. However, in resource-constrained settings (\textbf{Scenarios 1 and 2}), the TSP approach yields comparable performance due to limited optimization options under restricted BS resources.

Fig.\ref{Access_Prob_7b} compares the task completion probability $\eta$ when $N = 7$ BSs under four resource availability scenarios, following the same setup as Fig.\ref{Access_Prob_5b}. The proposed MCTS-based algorithm continues to outperform in resource-constrained environments (\textbf{Scenarios 1 and 2}). 
This performance gain is attributed to the expanded search space in denser BS networks, where the MCTS-based approach effectively utilizes its optimization flexibility to identify optimal BS-time pairs. As a result, the MCTS-based method achieves up to a $17\%$ higher $\eta$ compared to the TSP and $\epsilon$-greedy algorithms under resource-limited conditions. In contrast, under resource-abundant conditions (e.g., \textbf{Scenario 4}), the TSP approach shows better performance than the MCTS-based method due to its exhaustive BS visitation strategy. Visiting all 7 BSs nearly guarantees successful task offloading, however, it comes at the cost of significantly higher energy consumption. This trade-off is evident in Fig.~\ref{Power_Consum_integration}, where the TSP algorithm exhibits the highest energy usage among the compared methods.



Moreover, Fig.~\ref{Access_Prob_10b} illustrates $\eta$ when $N = 10$ BSs under four resource availability scenarios:
\textbf{Scenario 1}: 5 BSs with $>80\%$ spectrum availability, 5 BSs providing $50\%$ required CPU cycles
\textbf{Scenario 2}: 5 BSs with $>80\%$ spectrum availability, 7 BSs providing $50\%$ CPU cycles
\textbf{Scenario 3}: 7 BSs with $>80\%$ spectrum availability, 5 BSs offering large CPU cycles
\textbf{Scenario 4}: 7 BSs with high spectrum availability ($>80\%$) and large CPU cycles.

The $x$-axis represents the four scenarios, while the $y$-axis denotes the task completion probability $\eta$. Interestingly, while the proposed MCTS-based approach outperforms both benchmark methods across all scenarios, the performance improvement appears relatively smaller compared to the cases with $N = 5$ and $N = 7$ BSs. This is because when the system has sufficient resources, the cost of taking sub-optimal actions becomes less significant. In other words, the difference between visiting the BS with the highest spectrum availability and CPU cycles versus a sub-optimal BS with slightly lower resources is negligible, as the network as a whole can still satisfy the task offloading requirements. As a result, the eVTOL can successfully complete its offloading tasks regardless of the BS visiting sequence, and the main variation comes from the overall travel distance, which has limited impact on task completion performance in resource-rich environments. However, we would like to emphasize that while the improvement in $\eta$ appears limited, the reduction in energy consumption achieved by the proposed method is non-negligible, as detailed in Fig.~\ref{Power_Consum_integration}.




\begin{figure*}
\centering
\subfloat[ ]{\label{Reg_5}{\includegraphics[width=0.31\linewidth]{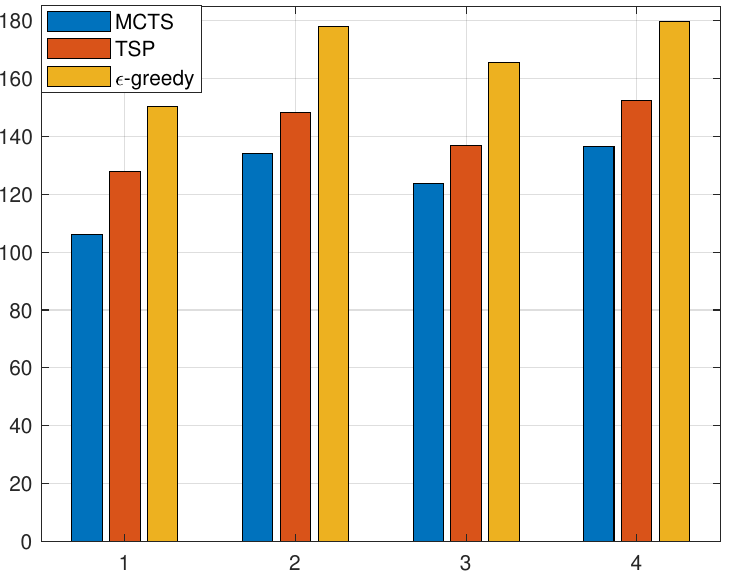}}} \hfill
\subfloat[ ]{\label{Reg_7}{\includegraphics[width=0.31\linewidth]{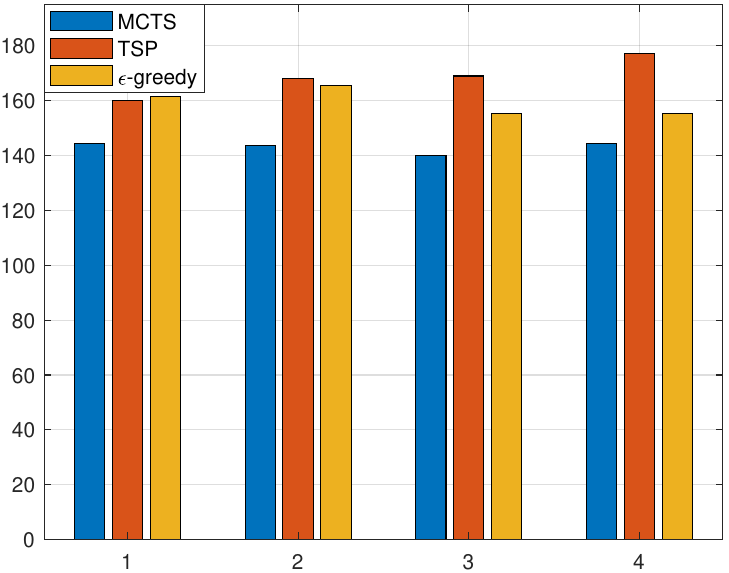}}} \hfill
\subfloat[ ]{\label{Reg_10}{\includegraphics[width=0.31\linewidth]{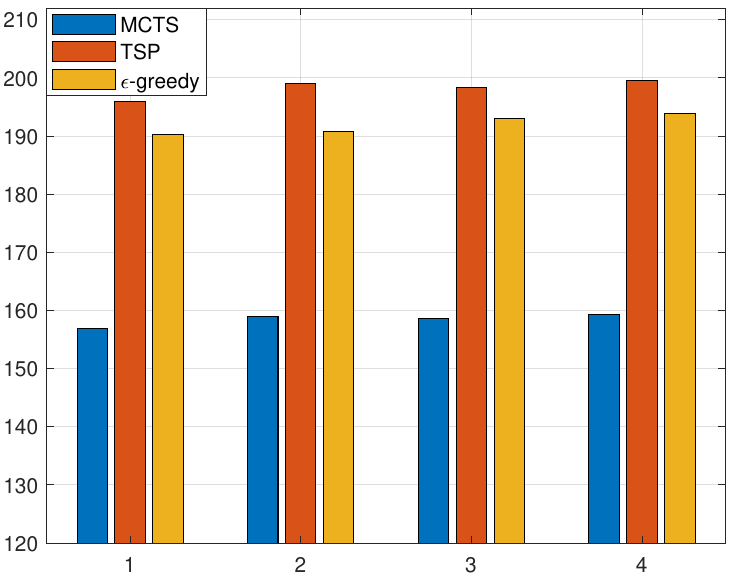}}} \hfill
\caption{Results of the objective function, $\mathcal{P}1$, under three algorithms when \textbf{(a)} $N = 5$ BSs, \textbf{(b)} $N = 7$ BSs, and \textbf{(c)} $N = 10$ BSs.}
\label{Objective_Fun_compare}
\end{figure*}

\subsection{Energy Consumption}
Fig.~\ref{Power_Consum_integration} compares the energy consumption of eVTOL operations, where the $x$-axis represents the number of accessed BSs, and the $y$-axis denotes the total energy consumption $\mathcal{W}_t$, defined as the sum of flight propulsion and hovering energy costs, as given in (\ref{eq_energyconstrains}).

In Fig.~\ref{5BS_PC}, we observe that when $N = 5$ BSs, the MCTS-based algorithm achieves the lowest energy consumption by completing the task offloading process through accessing only three BSs. In contrast, both the $\epsilon$-greedy and TSP algorithms require visiting all five BSs (i.e., full traversal). While the TSP approach exhibits lower energy consumption during the early stages, due to its shortest-path trajectory planning strategy, it ultimately results in higher overall energy usage once the complete task offloading is achieved.

Fig.~\ref{7BS_PC} presents the simulation results of energy consumption when $N = 7$ BSs. The proposed MCTS-based algorithm enables the eVTOL swarm to complete task offloading by accessing only two BSs. In contrast, the TSP approach exhibits linearly increasing energy consumption, as it requires the swarm to visit all BSs. The $\epsilon$-greedy algorithm completes the task at the fourth BS access but results in significantly higher energy consumption due to its suboptimal path planning compared to the MCTS-based method.

The energy efficiency advantage of the proposed MCTS-based approach becomes even more pronounced when $N = 10$ BSs. As the number of BSs increases, the MCTS-based algorithm consistently maintains the lowest energy consumption through intelligent BS selection, while the other two benchmark algorithms experience quadratic growth in energy usage, proportional to $N$. These results demonstrate the suitability of the proposed MCTS-based approach for large-scale AAM deployments.



\subsection{Objective Function}
Fig.~\ref{Objective_Fun_compare} presents the simulation results of the objective function $\mathcal{P}1$ under the three algorithms across different numbers of BSs. Recall that $\mathcal{P}1$ consists of two components: energy consumption (including both flight and hovering) and task failure probability (defined as the negative of task completion probability), obtained by summing the results from the previous two subsections.

By jointly considering both energy consumption and task completion performance, the proposed MCTS-based algorithm consistently outperforms the benchmark methods under all BS configurations ($N = 5, 7, 10$) and various resource availability scenarios. A smaller objective function value indicates better overall performance, as $\mathcal{P}1$ is defined as a minimization problem. Notably, the objective function value of the MCTS-based approach decreases as the number of BSs increases, further demonstrating its scalability and effectiveness in denser BS deployments.

\begin{figure*}
\centering
\subfloat[ ]{\label{Reg_5}{\includegraphics[width=0.31\linewidth]{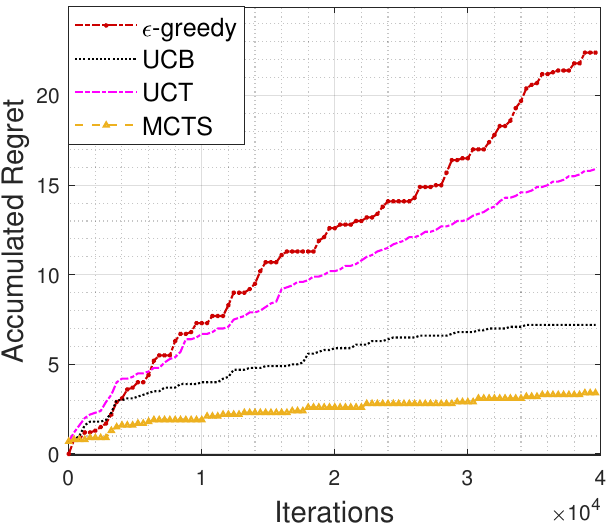}}} \hfill
\subfloat[ ]{\label{Reg_7}{\includegraphics[width=0.31\linewidth]{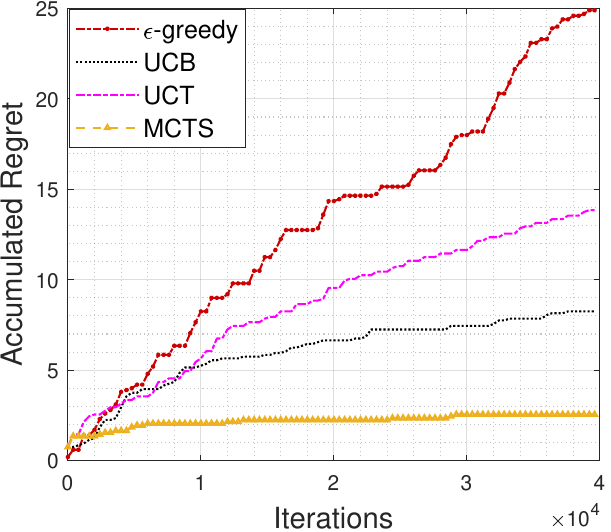}}} \hfill
\subfloat[ ]{\label{Reg_10}{\includegraphics[width=0.31\linewidth]{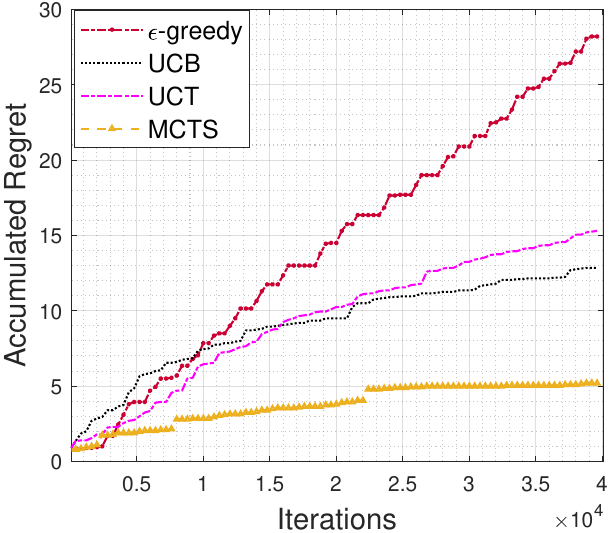}}} \hfill
\caption{Accumulative regrets under different algorithms when \textbf{(a)} $N = 5$ BSs, \textbf{(b)} $N = 7$ BSs, and \textbf{(c)} $N = 5$ BSs.}
\label{Regret_compare}
\end{figure*}

\begin{table}[!ht]
\centering
\caption{Successful Spectrum Access Probability}
\begin{tabular}{llll}
\hline
      & MCTS        & TSP           & $\epsilon$-greedy     \\ \hline
5-BS  & 88.24\% & 47\%    & 45.73\% \\
7-BS  & 88.83\% & 41.72\% & 86.64\% \\
10-BS & 93.3\% & 43\%   & 85.37\% \\ \hline
\end{tabular}
\label{successful_access_prob}
\end{table}

\begin{table}[!ht]
\centering
\caption{Average CPU Cycle Duration per Access}
\begin{tabular}{llll}
\hline
      & MCTS      & TSP      & $\epsilon$-greedy     \\ \hline
5-BS  & 6.33  & 2.8  & 2.8 \\
7-BS  & 15.5  & 4.43 & 7.75 \\
10-BS & 14.75 & 5.9  & 9.83 \\ \hline
\end{tabular}
\label{CPU_cycle_Duration}
\end{table}

\subsection{Spectrum access and CPU cycle}
Table~\ref{successful_access_prob} presents the successful spectrum access probabilities under the three algorithms across different numbers of BSs, considering \textbf{Scenario 1}, the resource-limited case defined earlier. The proposed MCTS-based algorithm demonstrates the highest robustness, achieving nearly 90\% access success rates across all scenarios. In contrast, the TSP algorithm exhibits the lowest access rates, while the $\epsilon$-greedy approach shows suboptimal performance, achieving 86\% in the $N = 7$ BSs configuration, still lower than that of the MCTS-based algorithm.

Table~\ref{CPU_cycle_Duration} summarizes the average CPU cycle duration per successful BS access. Notably, the MCTS-based approach outperforms the benchmark algorithms by obtaining significantly longer CPU cycle durations. Specifically, it achieves a 50\% increase compared to the $\epsilon$-greedy algorithm and a 150\% improvement over the TSP algorithm, highlighting its superior capability in selecting BSs with higher idle CPU availability during task offloading.


\subsection{Accumulated Regret Without Prior Estimation}
To further illustrate the convergence properties of the proposed algorithm, we plot the accumulated regret, as defined in (\ref{gfhfghdrt}). This metric is commonly used to quantify the divergence between algorithmic decisions and optimal actions. For this evaluation, we remove the prior estimation stage and require all algorithms to optimize BS and time selection through iterative trial-and-error processes, simulating a scenario in which the eVTOL swarm lacks prior knowledge of BS resource availability.


Fig.~\ref{Regret_compare} compares the cumulative regret of four algorithms:
(i) $\epsilon$-greedy, which probabilistically balances exploration and exploitation;
(ii) UCB, which optimizes using the upper confidence bound;
(iii) UCT, which integrates tree search without back-propagation \cite{Kocsis11871842_29};
and (iv) the proposed MCTS-based algorithm without prior estimation.

The $x$-axis represents the number of BS access attempts (iterations), and the $y$-axis shows the accumulated regret. As shown in Fig.\ref{Reg_5} (corresponding to Scenario 1 in Fig.\ref{Access_Prob_integration}), $\epsilon$-greedy exhibits linearly increasing regret due to its random exploration behavior. UCB and UCT show lower cumulative regrets as UCB dynamically balances the exploration–exploitation trade-off, while UCT leverages tree structures for decision-making. However, the lack of back-propagation in UCT hinders its ability to integrate historical experience, resulting in a final regret nearly 100\% higher than that of MCTS approach. In contrast, the MCTS algorithm achieves the fastest convergence by systematically updating the search tree through back-propagation.

\begin{figure*}
\centering
\subfloat[]{\label{Tra_5}{\includegraphics[width=0.31\linewidth]{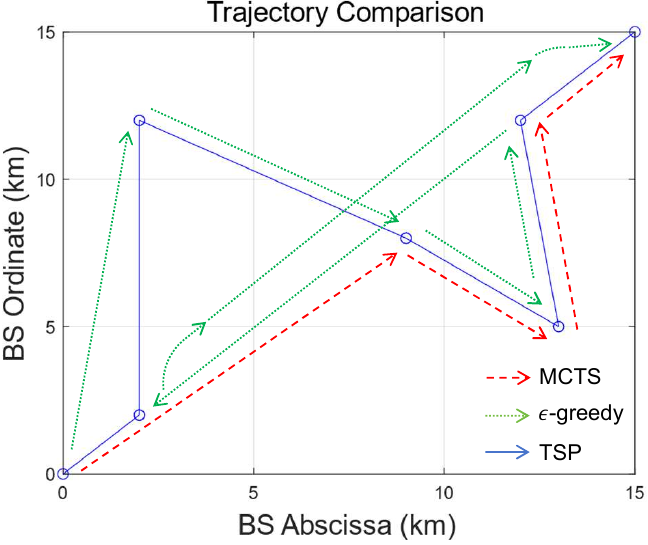}}} \hfill
\subfloat[]{\label{Tra_7}{\includegraphics[width=0.31\linewidth]{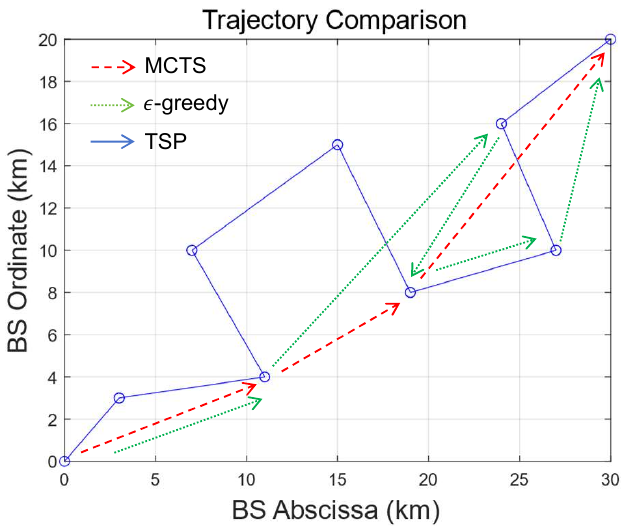}}} \hfill
\subfloat[]{\label{Tra_10}{\includegraphics[width=0.31\linewidth]{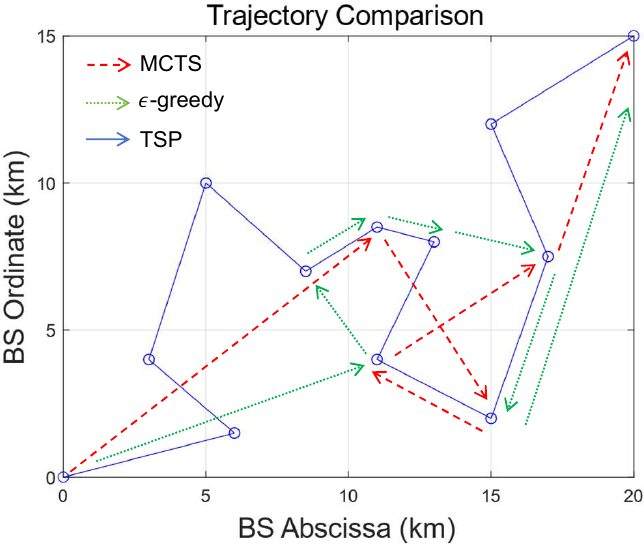}}} \hfill
\caption{Illustration of the eVTOL trajectories under different algorithms v.s. the number of BSs \textbf{(a)} $N = 5$ BSs, \textbf{(b)} $N = 7$ BSs, and \textbf{(c)} $N = 5$ BSs.}
\label{Trajectory_compare}
\end{figure*}


In Fig.\ref{Reg_7}, the $N = 7$ BSs case exhibits similar regret growth trends as the $N = 5$ BSs scenario. However, in the $N = 10$ BSs case (Fig.\ref{Reg_10}), UCB struggles to cope with the larger search space, with its regret increasing 1.8 times faster than in the $N = 5$ case due to excessive exploration. This comparison highlights the superior scalability of the proposed MCTS-based algorithm, which effectively limits regret growth despite the expanded state space, demonstrating its robustness in large-scale environments. As the search space grows, MCTS maintains faster convergence than the other algorithms. This advantage stems from its synergistic integration of a structured tree search architecture with experience-driven back-propagation, allowing for efficient optimization in high-dimensional resource allocation problems.

{\color{black} To further verify the robustness of the proposed method to different spectrum availability distributions (i.e., not being limited to the initial Gaussian assumption), we supplement a comparative simulation where spectrum availability follows a Poisson distribution. In this supplementary simulation, the parameter of the Poisson distribution is tuned to match the mean value of the Gaussian distribution used in the main simulations, ensuring comparable baseline resource availability.

\begin{figure} [h]
\centering
\includegraphics[width=0.65\columnwidth]{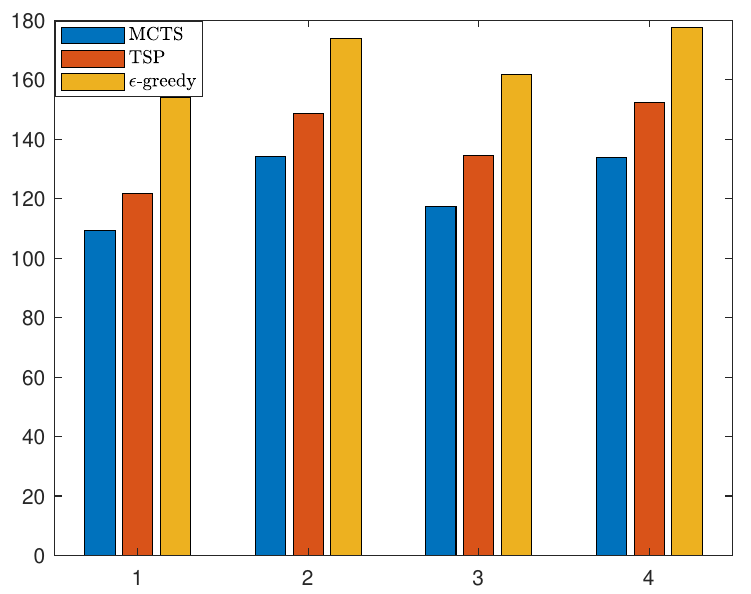} 
     \caption{Results of the objective function, $\mathcal{P}1$, $N=5$ BSs, under the Poisson distribution} 
\label{fig_10}
\end{figure}

Fig.~\ref{fig_10} illustrates the simulation results of $\eta$ under $N=5$ BSs and Poisson resource distribution.
The $x$-axis represents the four resource availability scenarios (that is same with Fig.~\ref{Access_Prob_integration}), while the $y$-axis shows the task completion probability $\eta$.

The results indicate that even when spectrum availability conforms to a Poisson distribution, the MAB-MCTS integration framework still achieves a high task completion rate and controls the increase in total energy consumption. These results are consistent with our theoretical analysis. The MAB-based estimation mechanism dynamically adapts to the statistical properties of spectrum availability (whether Gaussian or Poisson) via real-time interaction data, and the subsequent MCTS-driven trajectory and resource decision-making process remains effective. Thereby, this characteristic demonstrates the method's flexibility and robustness in different scenarios.}


\subsection{Flight Trajectory Comparison}

This subsection compares the flight trajectories generated by different algorithms under the $N = 5, 7, 10$ BS scenarios. Fig.~\ref{Trajectory_compare} illustrates the locations of BSs and the corresponding eVTOL flight paths for each case. The following key observations can be drawn:

For the TSP algorithm, the swarm is forced to exhaustively visit all BSs via the shortest-path routing, regardless of whether the task has already been completed. This leads to redundant flight paths and unnecessary energy expenditure. For instance, in the $N = 5$ BS scenario, the TSP trajectory results in a $20\%$ longer flight distance compared to the MCTS-based method, despite achieving the same task completion outcome.

In the $\epsilon$-greedy algorithm, the swarm is observed to revisit BSs they have previously accessed. This behavior arises from the stochastic nature of BS resource availability, failures in access or inadequate CPU cycles often cause the swarm to return to already-explored BSs. These erratic flight patterns increase energy consumption by up to $50\%$ compared to the MCTS-based algorithm in the $N = 5$ BS scenario.

In contrast, the proposed MCTS-based algorithm achieves task completion with the minimal number of BS accesses, thereby improving both energy and time efficiency. In the $N = 7$ BS scenario, the MCTS-based approach completes tasks with $71\%$ fewer BS visits than the TSP algorithm and $50\%$ fewer than the $\epsilon$-greedy approach, highlighting its superior resource-aware path optimization capability.



{\color{black}
\section{Conclusion}

This work proposes an MCTS-based trajectory optimization algorithm for eVTOL-involved CR networks, specifically designed to address the unique challenges of dynamic task offloading in AAM systems. By leveraging the spectrum-agile capabilities of CR, our framework first employs a MAB model to dynamically estimate the temporal availability of spectrum resources at terrestrial BSs. Subsequently, the eVTOL swarm utilizes a MCTS algorithm to determine the optimal sequence of BS visits and corresponding access time slots, jointly optimizing for task offloading success probability, energy consumption, and adherence to flight time constraints. The proposed approach also features an adaptive re-planning mechanism to handle access failures in real-time, ensuring robust task completion.
Numerical results demonstrate that the proposed MCTS-MAB framework significantly outperforms benchmark methods. It validates the effectiveness of our integrated spatio-temporal optimization strategy in managing the complex interplay between eVTOL mobility, dynamic BS resource availability, and stringent energy constraints.

While this work provides a foundational framework for intelligent trajectory and resource planning in CR-enabled AAM, it focuses on optimization for a single eVTOL swarm. A critical next step is to extend the framework to manage multiple, potentially competing swarms operating within the same airspace and CR spectrum.
Furthermore, to enhance the system's practicality, future research should integrate external factors into the eVTOL decision-making process. This includes no-fly zones, and real-time directives from Air Traffic Control (ATC) systems. Such integration would make the flight planning more adaptive and safety-compliant.

}

\bibliography{biblio}

\end{document}